\documentclass[aps,prl,twocolumn,showpacs,superscriptaddress,groupedaddress,preprintnumbers,bm,amsmath,amssymb,amsfonts,showpacs,superscriptaddress]{revtex4-1}

\usepackage[hyperindex,breaklinks]{hyperref}
\usepackage{graphicx,epstopdf}
\usepackage[utf8]{inputenc} 

\usepackage[usenames,dvipsnames]{color}
\usepackage{slashed}    

\def\be{\begin{equation}}
\def\ee{\end{equation}}
\def\ba{\begin{eqnarray}}
\def\ea{\end{eqnarray}}
\newcommand{\bea}{\begin{eqnarray}}
\newcommand{\eea}{\end{eqnarray}}

\newcommand{\ds}{{\sf DarkSUSY}}

\def\ge{\mathrel{\raise.3ex\hbox{$>$\kern-.75em\lower1ex\hbox{$\sim$}}}}
\def\la{\mathrel{\raise.3ex\hbox{$<$\kern-.75em\lower1ex\hbox{$\sim$}}}}

\def\simgt{\mathrel{\raise.3ex\hbox{$>$\kern-.75em\lower1ex\hbox{$\sim$}}}}
\def\simlt{\mathrel{\raise.3ex\hbox{$<$\kern-.75em\lower1ex\hbox{$\sim$}}}}

\newcommand{\nc}{\newcommand}

\nc{\gone}{\bar g_{\pi NN}^{(1)}}
\nc{\gzero}{\bar g_{\pi NN}^{(0)}}
\nc{\al}{\alpha}
\nc{\ga}{\gamma}
\nc{\de}{\delta}
\nc{\ep}{\epsilon}
\nc{\ze}{\zeta}
\nc{\et}{\eta}
\nc{\ka}{\kappa}
\nc{\rh}{\rho}
\nc{\si}{\sigma}
\nc{\ta}{\tau}
\nc{\up}{\upsilon}
\nc{\ph}{\phi}
\nc{\ch}{\chi}
\nc{\ps}{\psi}
\nc{\om}{\omega}
\nc{\Ga}{\Gamma}
\nc{\De}{\Delta}
\nc{\La}{\Lambda}
\nc{\Up}{\Upsilon}
\nc{\Ph}{\Phi}
\nc{\Ps}{\Psi}
\nc{\Om}{\Omega}
\nc{\ptl}{\partial}
\nc{\del}{\nabla}
\nc{\ov}{\overline}
\nc{\newcaption}[1]{\centerline{\parbox{15cm}{\caption{#1}}}}
\nc{\us}{U(1)$_S$}

\def\beq{\begin{equation}}
\def\eeq{\end{equation}}
\def\bmat{\begin{displaymath}}
\def\emat{\end{displaymath}}
\def\bear{\begin{eqnarray}}
\def\eear{\end{eqnarray}}
\def\ba{\begin{eqnarray}}
\def\ea{\end{eqnarray}}
\def\bery{\begin{array}}
\def\ery{\end{array}}
\def\bit{\begin{itemize}}
\def\eit{\end{itemize}}
\def\ben{\begin{enumerate}}
\def\een{\end{enumerate}}
\def\btab{\begin{tabular}}
\def\etab{\end{tabular}}
\def\btbl{\begin{table}}
\def\etbl{\end{table}}
\def\bfig{\begin{figure}[htb]}
\def\efig{\end{figure}}
\def\bpic{\begin{picture}}
\def\epic{\end{picture}}

\def\ga{\mathrel{\raise.3ex\hbox{$>$\kern-.75em\lower1ex\hbox{$\sim$}}}}
\def\la{\mathrel{\raise.3ex\hbox{$<$\kern-.75em\lower1ex\hbox{$\sim$}}}}
\def\gappeq{\mathrel{\rlap {\raise.5ex\hbox{$>$}}
{\lower.5ex\hbox{$\sim$}}}}
\def\lappeq{\mathrel{\rlap{\raise.5ex\hbox{$<$}}
{\lower.5ex\hbox{$\sim$}}}}

\def\gyr{{\rm \, G\kern-0.125em yr}}
\def\mev{{\rm \, Me\kern-0.125em V}}
\def\gev{{\rm \, Ge\kern-0.125em V}}
\def\tev{{\rm \, Te\kern-0.125em V}}

\begin{document}
 

\title{Novel direct detection constraints on light dark matter}

\author{Torsten Bringmann}
\affiliation{Department of Physics, University of Oslo, Box 1048, N-0371 Oslo, Norway}

\author{Maxim Pospelov}
\affiliation{Perimeter Institute for Theoretical Physics, Waterloo, ON N2J 2W9, 
Canada}
\affiliation{Department of Physics and Astronomy, University of Victoria, 
Victoria, BC V8P 5C2, Canada}

\date{June 2018}

\begin{abstract}
\noindent 
All attempts to directly detect particle dark matter (DM) scattering on nuclei suffer from the partial or 
total loss of sensitivity for DM masses in the GeV range or below. We derive 
novel constraints from the inevitable existence of a subdominant, but highly energetic, component 
of DM generated through collisions with cosmic rays. 
Subsequent scattering inside conventional DM detectors, as well as neutrino 
detectors sensitive to nuclear recoils, limits the
DM-nucleon scattering cross section to be below $10^{-31}$\,cm$^2$
for both spin-independent and spin-dependent scattering of light DM.
\end{abstract}
\maketitle

\paragraph{Introduction.---} 

Attempts to discover non-gravitational interactions of dark matter (DM) are a global effort, 
pursuing many possible avenues -- 
perhaps as many as there are viable microscopic models that link DM with the rest of fundamental 
physics~\cite{Bertone:2004pz,Feng:2010gw}. The simplicity of the early Universe suggests that DM
may be realized in the form of some relic particles~\cite{Lee:1977ua,Gondolo:1990dk}, remnants 
of the Big Bang, that we denote here as $\chi$.

Among the very few things known about the galactic component of DM is the scale of its 
velocity, \mbox{$v_{\chi, {\rm gal}}\sim 10^{-3}c$}.
The energy carried by DM particles,
$E_{\chi} \sim m_{\chi} v_{\chi, {\rm gal}}^2$, can be shared with an atomic nucleus in the 
process of a collision, and therefore in principle be detected~\cite{Goodman:1984dc}.
The search for such DM-nucleus scatterings -- commonly referred to
as direct DM detection -- has seen several generations 
of experiments with ever improving sensitivity. In the absence of a credible positive signal,
this has translated to continuously tightening limits. The latest results from the XENON1T 
collaboration~\cite{Aprile:2018dbl} 
bring the sensitivity to the cross section {\it per nucleon} below the 
$\sigma_{\chi} = 10^{-46}\,$cm$^2$ level for 
the ``optimal" DM mass range, $m_\chi \in 15-100$\,GeV. This significantly 
constrains many models of weak-scale DM (see, {\em e.g.}~\cite{Athron:2017yua}). 

Below that mass range, and especially below 1\,GeV, the direct sensitivity to DM worsens rapidly. 
This is because the nuclear recoil energy becomes smaller, and cannot exceed 
$E_{\rm recoil}^{\rm max} = 2 m_\chi^2(v_{\rm esc})^2/m_{A}$, where $v_{\rm esc}\sim540$\,km/s is the galactic 
escape velocity and $m_{ A}$ the nuclear mass. If $E_{\rm recoil}^{\rm max}$ is below some 
detector threshold $E_{\rm thr}$, the sensitivity completely disappears, making even cross sections 
parametrically larger than weak-scale cross sections ({\em e.g.}~$\sigma \gg 10^{-36}\,{\rm cm}^2$) 
completely undetectable. 

Recently, it has been realized that several physical processes allow to circumvent this limitation.
If the scattering on the nucleus results in the emission of a photon or ejection 
of an atomic electron, {\it e.g.}, the electromagnetic fraction of the deposited energy can 
be larger than for elastic nuclear recoils, improving the sensitivity 
for $m_\chi$ in the few 100\,MeV range \cite{Kouvaris:2016afs,Ibe:2017yqa,Dolan:2017xbu}. 
Further constraints derive from {\em multiple} collisions of light DM. 
For example, interactions with 
fast moving nuclei or electrons inside the Sun can accelerate the DM above 
threshold for direct detection \cite{Kouvaris:2015nsa,An:2017ojc,Emken:2017hnp}. This 
contribution typically does not exceed a fraction of $\mathcal{O}(10^{-5})$ times the total DM flux on 
Earth, but nevertheless greatly enhances the mass reach of existing detectors, especially 
for $\chi-e^-$ scattering \cite{An:2017ojc}.

In this Letter, we consider another inevitable component of the DM flux, with velocities much 
higher than $v_{\rm esc}$. It originates from energetic galactic cosmic rays (CRs) 
colliding with cold DM particles in the Milky Way halo,
creating a secondary DM component of CRs with \mbox{(semi-)}relativistic momenta. 
This new component of the DM flux, called CRDM throughout this work, will scatter 
again in the detectors, but now with 
much greater energy available.
The goal of this work is to make use of this idea, employing data from the most sensitive current direct detection {\em and} 
neutrino experiments, to establish new direct limits on DM-nucleon scattering that extend to 
small DM masses (formally even to $m_\chi \to 0$).

We will adopt a simple two parameter model, $\{m_\chi,\sigma_{\chi }\}$, 
without reference to a specific underlying theory. For the DM-nucleon elastic cross section we assume for 
simplicity the isospin-singlet structure, $\sigma_{\chi n}=\sigma_{\chi p}\equiv \sigma_{\chi }$, but 
will consider both spin-dependent and spin-independent scattering.
DM  models with light $m_\chi$ often require sub-electroweak scale 
mediators \cite{Boehm:2003hm,Pospelov:2007mp}, and therefore can be amenable to 
additional constraints from cosmology, colliders, neutrino and beam dump experiments 
(see {\em e.g.}~Ref.~\cite{Battaglieri:2016ggd} for a review). 
However, all such constraints are necessarily 
model-dependent, while constraints derived in this Letter 
have  greater generality.
Despite invoking DM-CR interactions, in particular, they build on the same microscopic
picture of DM-nucleon scattering as adopted in the standard presentation of
limits from conventional direct detection experiments. 

\smallskip
\paragraph{Step 1: From CR to DM fluxes.--- } 
Compared to CR velocities, DM can be considered effectively at rest. Then, 
the kinetic energy transferred to a DM particle in a single collision is 
\bea
\label{eq:Tr}
T_\chi \!=T_\chi^\mathrm{max}\frac{1 \!-\! \cos\theta}{2}\,, ~
T_\chi^\mathrm{max}\!=\frac{T_i^2+2m_iT_i}{T_i \!+\! (m_i\!+\!m_\chi)^2\!/(2m_\chi\!)}\,,~~~
\eea
where $\theta$ is the c.m.s.~scattering angle and $T_i\equiv E_i-m_i $ the 
kinetic energy of the incoming CR particle $i$.
The (space-like) momentum transfer in the collision is given by $Q^2 = 2m_\chi T_\chi$.
For {\it isotropic } CR-DM scattering, both $T_\chi$ and $Q^2$ thus follow a flat 
distribution, with $T_\chi$ ranging from $0$ to $T_\chi^\mathrm{max}$. 
Inverting Eq.~(\ref{eq:Tr}) gives the {\it minimal} incoming CR 
energy required to obtain a DM recoil energy $T_\chi$:
\be
\label{eq:Tmin}
T_i^\mathrm{min}=\left( \frac{T_\chi}{2}  \!-\! m_i\right) \left[
1\pm\sqrt{1+\frac{2 T_\chi}{m_\chi}\frac{\left(m_i \!+\! m_\chi\right)^2}{\left(2m_i \!-\! {T_\chi}\right)^{2}}}
\right],
\ee
where the $+$ ($-$) sign applies for $T_\chi>2m_i$ ($T_\chi<2m_i$). 

The local interstellar (LIS) population of CRs is well measured and typically 
described by their {\it differential intensity} $dI/dR$, where $R$ is the particle's 
rigidity. 
We adopt parameterizations~\cite{DellaTorre:2016jjf,Boschini:2017fxq} 
for $dI_i/dR_i$ of protons and $^4$He nuclei, the two dominant CR components.
The {\it differential CR flux} (number of particles per area, kinetic energy and time) 
is then obtained as ${d\Phi}/{dT }=4\pi \left({dR}/{dT}\right) \left({dI}/{dR}\right)$.
For an  elastic scattering cross section $\sigma_{\chi i}$,  the collision rate 
of CR particles $i$ with energy in the range $[T_{i}, T_{i}+dT_{i}]$
inside a volume $dV$  thus becomes
\be
  d\Gamma_{{\rm CR}_i\to \chi} = \sigma_{\chi i} \times   \frac{\rho_{\chi}}{m_{\chi}} \frac{d\Phi^{LIS}_{i}}{dT_{i}} dT_{i} dV\,.
\ee
The resulting CR-induced DM flux is thus obtained by dividing by $4\pi d^2$, where $d$ is the
distance to the source, implying that the volume integration 
reduces to an angular average over a line-of-sight integral:
\bea
\frac{d\Phi_{\chi}}{dT_{i}}\! =\! \int\!\frac{d\Omega}{4\pi} \!\int_{l.o.s.} \!\!\!\!\!\!\!d\ell \,  \sigma_{\chi i} \frac{\rho_\chi}{m_\chi} \frac{d\Phi_{i}}{dT_{ i}}  
\equiv  \sigma_{\chi i } \frac{\rho_{\chi}^\mathrm{local}}{m_{\chi}} \frac{d\Phi^{LIS}_{i}}{dT_{i}} D_\mathrm{eff}\,.~~~
\label{eq:chiCR}
\eea
Here we introduced an {\it effective distance} $D_\mathrm{eff}$ out to which we take into account  CRs as the 
source of a possible high-velocity tail in the DM velocity distribution. 
The local gradient in the DM density is relatively well constrained~\cite{Catena:2009mf,Benito:2019ngh}, and that in the 
cosmic-ray density is very
small~\cite{Evoli:2012ha}. The main uncertainty in $D_\mathrm{eff}$ thus derives from the extension of the 
diffusion zone, which reaches out to at least a few kpc from the galactic 
disk~\cite{Evoli:2008dv,Moskalenko:2001qm,Bringmann:2011py}.
Assuming an NFW 
profile~\cite{Navarro:1995iw, Ackermann:2012rg} for the DM distribution
and a homogeneous CR distribution, e.g., performing the full line-of-sight integration out to 1\,kpc (10\,kpc) 
results in $D_\mathrm{eff}=0.997\,$kpc ($D_\mathrm{eff}=8.02\,$kpc). 
Using Eq.~(\ref{eq:Tr}), we can finally express the DM flux in terms of the DM energy by
integrating over all CR energies $T_i$: 
\be
\frac{d\Phi_{\chi}}{dT_{\chi}} = \int_0^\infty dT_{i} \frac{d\Phi_{\chi}}{dT_{i}} \frac{1}{T_{\chi}^\mathrm{max}(T_{i})} 
\Theta\left[T_{\chi}^\mathrm{max}(T_{i})-T_{\chi}\right]\,.
\ee

The flat distribution over recoil energies that follows from Eq.~(\ref{eq:Tr}) for isotropic scattering is an assumption 
that we modify by the inclusion of the hadronic elastic scattering form-factor in the simplest dipole form \cite{Perdrisat:2006hj}, 
\be
   G_i(Q^2) = 1/ (1 + Q^2 /\Lambda_i^2 )^2\,.
\ee 
Here, $\Lambda_i$ scales inversely proportional with the charge radius and is hence smaller for heavier nuclei;
for proton (Helium) scattering due to a vector current, one has $\Lambda_p \simeq 770$\,MeV 
($\Lambda_{\mathrm{He}} \simeq 410$\,MeV)~
\cite{Angeli:2004kvy}). 
 We thus relate the scattering cross section to that in the point-like limit by 
\be
  \frac{d\sigma_{\chi i}}{d\Omega} = \left.\frac{d\sigma_{\chi i}}{d\Omega} \right |_{Q^2=0} G_i^2(2m_{\chi} T_{\chi})\,.
\ee

\begin{figure}[t]
 \includegraphics[width=\columnwidth]{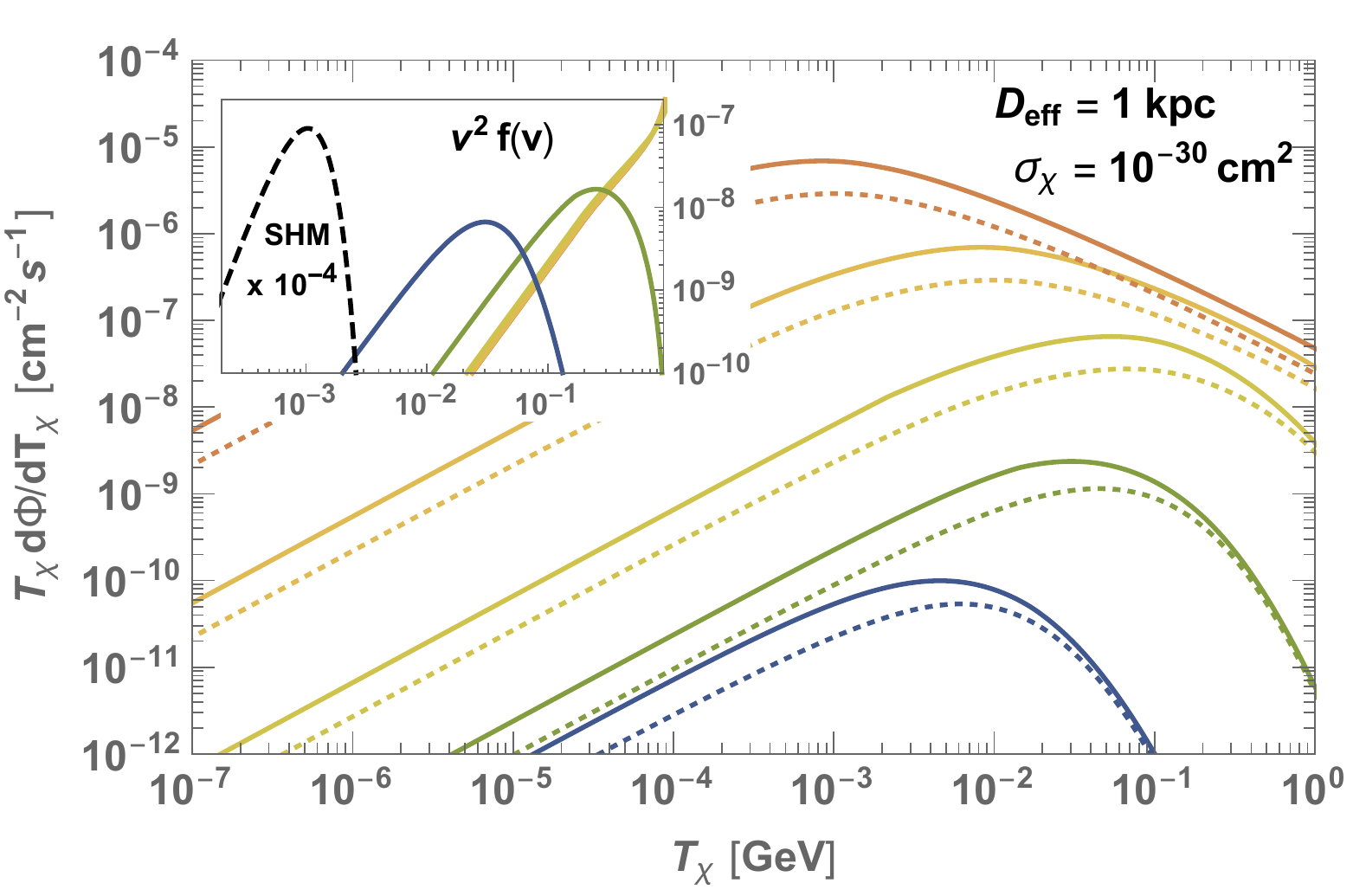}
 \caption{%
 Expected flux of CRDM for different DM masses $m_\chi=0.001,0.01,0.1,1,10$\,GeV (from top to bottom). 
 Dotted lines show the contribution from CR proton scattering
 alone. The flux is directly proportional to the effective distance $D_{\rm eff}$ and the elastic scattering cross 
 section $\sigma_\chi$, chosen here as \mbox{indicated.}
The inset compares the corresponding 1D velocity distributions
 $f(v)$, in units where $c=1$, to that of the standard halo model (dashed line).
  }
  \vspace*{-0.3cm}
 \label{CRDMflux}
\end{figure}

Putting everything together, we expect the following CR-induced DM flux:
\bea
\label{eq:dPhiz}
\frac{d\Phi_{\chi}}{dT_{\chi}} &=&  D_\mathrm{eff} \frac{\rho_{\chi}^\mathrm{local}}{m_{\chi}}  \times\\
&& \times \sum_i \sigma^0_{\chi i}\, G_i^2(2m_{\chi} T_{\chi}) 
\int_{T_{i}^\mathrm{min}}^\infty d T_{i}\, \frac{{d\Phi^{LIS}_{i}}/{dT_{i}}}{T_{\chi}^\mathrm{max}(T_{i})} \,.\nonumber
\eea
Here, we only include $i\in \{p,\,^4\mathrm{He}\}$  in the sum.
In Fig.~\ref{CRDMflux} we plot these CRDM fluxes for various DM masses,
for spin-independent $\sigma_\chi=\sigma_n=\sigma_p$. The 
contribution from Helium can be even  larger than that from protons, but is
formfactor-suppressed at large recoil energies. The flux is related to the 1D velocity distribution 
$f(v)$, more familiar in the context of direct DM searches, as 
$f(v)=m_\chi^2({\rho_{\chi}^\mathrm{local}})^{-1}\gamma^3{d\Phi_{\chi}}/{dT_{\chi}}$.
For illustration, we compare this to the Maxwellian distribution of the standard 
halo model \cite{Drukier:1986tm}, displayed as a dashed line in the inset. As expected, the CRDM 
population peaks at \mbox{(semi-)}relativistic velocities, and is highly
subdominant at the galactic DM velocities typically considered. 

\smallskip
\paragraph{Step 2: Attenuation of CRDM flux.---}
Very large scattering cross sections generally constitute a blind spot for direct DM  detection,
because they would lead to a significant attenuation of the DM flux from the top of the 
atmosphere to the location of the 
detector~\cite{Starkman:1990nj,Mack:2007xj,Hooper:2018bfw,Emken:2018run}.
The degradation in energy should also 
occur for the CRDM component,
and we can estimate  the energy loss of DM particles propagating through 
a medium as
\be
\frac{dT_{DM}}{dx}=
-\sum_N n_N
\int_0^{T_r^\mathrm{max}} \frac{d\sigma_{\chi N}}{dT_r} T_r d T_r\,.
\label{eq:eloss}
\ee
Here, $T_r$ refers to the energy lost by a CRDM particle in a collision with 
nucleus $N$. This process, in analogy with neutrino scattering, can be elastic, quasi-elastic 
or inelastic. The latter two are likely to dominate at high energies $T_\chi > {\rm few\,100\, MeV}$. 
(In a quasi-elastic process one or more nucleons are dislodged from $N$, while in an inelastic process 
additional hadrons are created in the $\chi-N$ collision.) In this work we will limit ourselves to 
elastic scattering, leaving a more elaborate treatment for future considerations. Using the uniform 
distribution of the nuclear recoil energy for isotropic scattering, we have 
$d\sigma_N/dT_r=\sigma_N/T_r^\mathrm{max}$,
and hence
\bea
\nonumber
\frac{dT_{\chi}}{dx}&=&-\frac{1}{2}
\sum_N n_N \sigma_{\chi N} T_r^\mathrm{max}\approx
-\frac{1}{2m_{\chi}\ell}  \left(T_{\chi}^2+2m_{\chi}T_{\chi}\right)\,,\\
&&{\rm where }~~\ell^{-1} \equiv \sum_N n_N \sigma_{\chi N} \frac{2 m_N m_{\chi}}{(m_N+m_{\chi})^2} \,.
\label{eq:ldef}
\eea
In the last step we have assumed $T_{\chi}\ll m_N$ in Eq.~(\ref{eq:Tr}). 
Integrating this equation, we can relate, very approximately, the differential DM 
flux at depth $z$ to the one at the top of the atmosphere as
\be
{\frac{d\Phi_{\chi}}{dT_{\chi}^z} = \left(\frac{dT_{\chi}}{dT_{\chi}^z}\right) \frac{d\Phi_{\chi}}{dT_{\chi}}
= \frac{4 m_{\chi}^2e^{z/\ell}}{\left(  
2m_{\chi}+{T_{\chi}^z}  -{T_{\chi}^z} e^{z/\ell} 
\right)^{2}}  \frac{d\Phi_{\chi}}{dT_{\chi}}}\,,
\ee 
where ${d\Phi_{\chi}}/{dT_{\chi}}$, needs to be evaluated at 
\be
\label{eq:TztoT}
\!\!\!\! T_{\chi}=T_{\chi}^0(T_{\chi}^z)=
2m_{\chi} T_{\chi}^ze^{z/\ell}\left(  
2m_{\chi} \!+\! {T_{\chi}^z}  \!-\! {T_{\chi}^z} e^{z/\ell} 
\right)^{-1}.
\ee
For ${T_{\chi}^0}\ll{m_{\chi}}$ our treatment of the energy attenuation reduces to 
that previously considered in Ref. \cite{Emken:2018run}.

For the mean free path of the DM particles, $\ell$,  we 
use \ds~\cite{Bringmann:2018lay} to calculate the {average} density
$n_N$ of Earth's 11 most abundant elements between surface and
depth $z$ (based on mass density profiles from Ref.~\cite{2003TrGeo...2..547M}).
We also need to relate the nuclear cross sections to the one on 
nucleons, $\sigma_\chi$. For spin-independent scattering, there is the usual coherent enhancement,
leading to 
\be
\label{eq:chiN}
  \sigma_{\chi N }= \sigma_{\chi}^\mathrm{SI} A^2 \left(\frac{m_N(m_\chi+m_p)}{m_p(m_\chi+m_N)}\right)^2\,.
\ee
We neglect nuclear form-factors  in obtaining $\ell$. Along with the energy-loss 
ansatz (\ref{eq:eloss}), as compared to full numerical 
simulations~\cite{Emken:2018run},
this leads to  conservative limits.

\smallskip
\paragraph{Step 3: CRDM scattering in detectors.---} 
Once a CRDM particle reaches a detector at depth $z$, it can transfer (part of its) energy 
to a target nucleus inside the detector. Exploiting completely analogous formulae to the case of DM$\to$CR
scattering discussed above, in particular the flat distribution of the target nucleus recoil energy $T_N$  
for a given DM energy, we find the differential recoil rate {\it per target nucleus} 
to be
\be
\label{eq:DDscat}
\frac{d\Gamma_N}{d T_{N}}= \sigma_{\chi N}^0  G_N^2(2m_NT_{N}) \   \int_{T_{\chi}(T_{\chi}^{z, \mathrm{min}})}^\infty \!\!\ 
  \frac{ dT_{\chi}}{T_{r,N}^{\mathrm{ max}}(T_\chi^z)}\frac{d\Phi_{\chi}}{dT_{\chi}}\,.
\ee
Here $G_N(Q^2)$ is a nuclear form-factor and ${d\Phi_{\chi}}/{dT_{\chi}}$ is given in Eq.~(\ref{eq:dPhiz});
the quantities $T_{r,N}^{\mathrm{ max}}$ and $T_{\chi}^{z, \mathrm{min}}$ follow from 
Eqs.~(\ref{eq:Tr}) and (\ref{eq:Tmin}), 
by replacing $\chi\to N$ and $i\to\chi$.

The broad energy distribution of CRDM particles allows us, based on Eq.~(\ref{eq:DDscat}), to use 
both conventional direct detection and neutrino experiments to set novel limits on $\sigma_{\chi}$.
It is clear that for small enough $\sigma_\chi$ the overburden mass above the detectors is transparent 
to CRDM, and the overall strength of the signal hence scales as $\sigma_\chi^2$. 
For large cross sections, on the other hand, the strong attenuation of the CRDM energy as 
given in Eq.~(\ref{eq:TztoT}) also leads to an exponential suppression of the signal. 

\smallskip
\paragraph{Resulting limits.---}
We begin by addressing constraints from conventional direct detection experiments,
which we derive from reported limiting values for heavy DM cross sections on nucleons as a function of 
the DM particle mass, ${\sigma_\mathrm{DM}^\mathrm{SI,lim}}(m_{\rm DM})$.
Assuming a non-relativistic DM velocity distribution $f_\mathrm{NR}(v)$,
and hence a standard DM flux of 
$d\Phi_\mathrm{DM}/dT_\mathrm{DM}=m_\mathrm{DM}^{-2}\,{\rho_\mathrm{DM}^\mathrm{local}}f_\mathrm{NR}$,
we can relate the count rate per target nucleus $N$ to the average heavy DM-nucleus cross section 
$\sigma_{\chi N}^\mathrm{DM}$ inside  the experimentally accessible window of recoil energies $T_N\in\{T_1,T_2\}$.  
{\it In the limit of large DM masses}, this gives
\bea
\Gamma_N^\mathrm{DM}  &=& \!\int_{T_1}^{T_2}\!\!\!dT_N\, \sigma_{\chi N}^\mathrm{DM} \int_0^\infty\!\!\!\!\!
 dT_\mathrm{DM} \frac{d\Phi_\mathrm{DM}}{dT_\mathrm{DM}} 
\frac{\Theta\left[T_N^\mathrm{max}(T_\mathrm{DM}) \!-\! T_N\right]}
{T_N^\mathrm{max}(T_\mathrm{DM})} \nonumber\\
&\simeq& \kappa \frac{\sigma_{\chi N}^\mathrm{DM}}{m_\mathrm{DM}} 
~(\bar v\, \rho_\mathrm{DM})^\mathrm{local} \quad \mathrm{for~} m_\mathrm{DM}\gg m_N
\,, \label{eq:GammaN}
\eea
Here $\bar v$ denotes the mean DM velocity and $\kappa$ is an $\mathcal{O}(1)$ constant that,
for a Maxwellian distribution, equals 
$\kappa=\exp[-2 T_1/(\pi m_N \bar v^2)]-\exp[-2 T_2/(\pi m_N \bar v^2)]$.

\begin{figure}
 \mbox{ }\vspace*{-0.28cm}\\
 \includegraphics[width=\columnwidth]{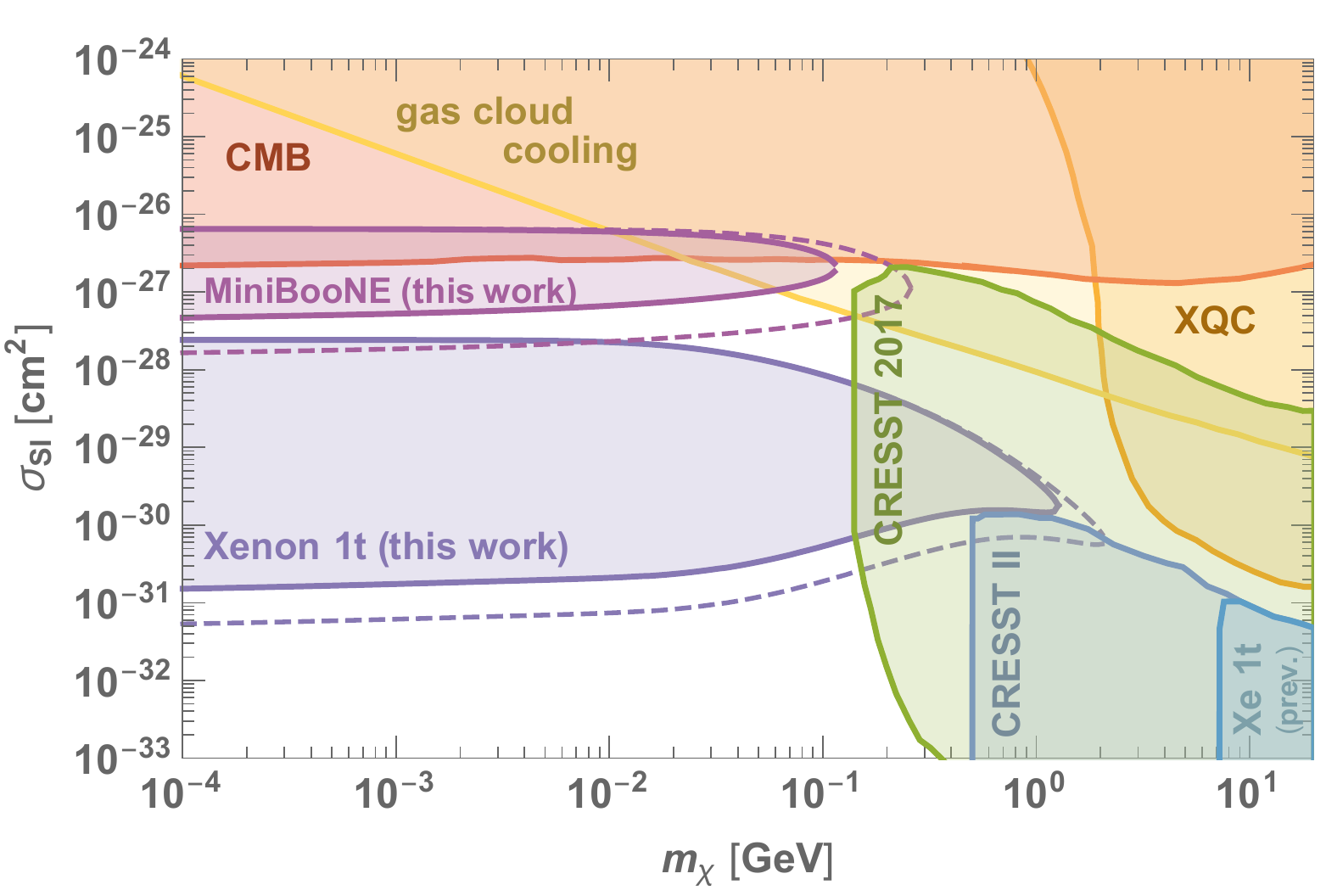}
 \caption{%
 Constraints on spin-independent DM-nucleon scattering imposed by the XENON-1T and MiniBooNE experiments.
 Solid (dashed) lines assume a CR density that equals, on average, the local value out to a distance of
 1\,kpc (10\,kpc). We compare our limits to those deriving from CMB observations~\cite{Xu:2018efh}, 
 gas cloud cooling~\cite{Bhoonah:2018wmw}, the X-ray Quantum Calorimeter experiment (XQC) \cite{Mahdawi:2018euy}, and
 a selection of direct detection 
 experiments~\cite{Angloher:2015ewa,Angloher:2017sxg,Aprile:2017iyp} 
 after taking into account the absorption of DM in soil and atmosphere~\cite{Emken:2018run}. 
 \label{constraintsSI}
}
\end{figure}

In order to constrain the CRDM component we now need to compare Eq.~(\ref{eq:GammaN}) with 
Eq.~(\ref{eq:DDscat}), taking into account that $\sigma_{\chi N}^0$ is evaluated for 
$m_\mathrm{DM}\gg m_N$ only in the former case. For spin-independent scattering, we can 
use Eq.~(\ref{eq:chiN}) to compute the ratio of these cross sections. Realizing that the coherence 
factors for $\sigma_{\chi N}$ are identical between ordinary DM and CRDM scattering, then 
allows us to recast conventional limits on the scattering rate $\sigma_\mathrm{DM}^\mathrm{SI, lim}$ 
{\it per nucleon} 
to an equivalent limit resulting from the CRDM component:
\bea
\sigma_{\chi}^\mathrm{SI.lim}&=& \kappa\, (\bar v\, \rho_\mathrm{DM})^\mathrm{local}
\left(\frac{m_\chi+m_N}{m_\chi+m_p}\right)^2
 \bigg(\frac{\sigma_\mathrm{DM}^\mathrm{SI, lim}}{m_\mathrm{DM}} \bigg)_{m_\mathrm{DM}\to\infty}\nonumber\\
 &&\times\bigg(
  \int_{T_1}^{T_2}dT_N\int_{T_{\chi}(T_{\chi}^{z, \mathrm{min}})}^\infty \!\!\ 
  \frac{ dT_{\chi}}{T_{r,N}^{\mathrm{ max}}}\frac{d\Phi_{\chi}}{dT_{\chi}}
 \bigg)^{-1}
\eea
For the recent Xenon 1T data (Fig. 5 of \cite{Aprile:2018dbl}), e.g., one has 
${\sigma_\mathrm{DM}^\mathrm{SI,lim}}/{m_\mathrm{WIMP}}= 8.3\cdot10^{-49}$\,cm$^2$/GeV 
for $m_\chi\gtrsim100$\,GeV, and $T_\mathrm{Xe}\in[4.9,40.9]$\,keV 
implies $\kappa\simeq0.23$. 
The resulting limits on $\sigma_\chi$ are
shown in Fig.~\ref{constraintsSI}, for different assumptions about the size of the diffusion zone
(with solid lines corresponding to an ultra-conservative choice). 
For small DM masses these limits exclude cross sections in the range 
$10^{-31} \,{\rm cm}^2\lesssim \sigma_{\chi}^\mathrm{SI}  \lesssim10^{-28}\,{\rm cm}^2$,
almost independently of $m_\chi$. 
Clearly, these constraints are highly complementary to existing limits on light 
DM~\cite{Erickcek:2007jv,Emken:2018run,Gluscevic:2017ywp,Bhoonah:2018wmw,Mahdawi:2018euy,Xu:2018efh,Collar:2018ydf}. 
Direct detection of light energetic dark sector particles was also discussed in Ref.~\cite{Cui:2017ytb}.

Due to its shallow location, MiniBooNE~\cite{Karagiorgi:2006jf}  gives a particular advantage in 
limiting CRDM fluxes with large scattering cross sections that prevent $\chi$ from reaching  
deeply placed experiments. We utilize the measurement of elastic $\nu-p$ 
scattering~\cite{Aguilar-Arevalo:2013nkf}, and a recent DM run~\cite{Aguilar-Arevalo:2017mqx} 
that allows to extract the beam-unrelated scattering rate. Requiring the  
scattering rate of CRDM on protons at MiniBooNE depth not to exceed the 
beam-unrelated background, we obtain (see Appendix)
\be
\Gamma_p (T_p>35\,{\rm MeV})< 1.5\times 10^{-32}\,{\rm s}^{-1}.
\ee
This additional exclusion region is also shown in Fig.~\ref{constraintsSI}. 

 \begin{figure}
 \includegraphics[width=\columnwidth]{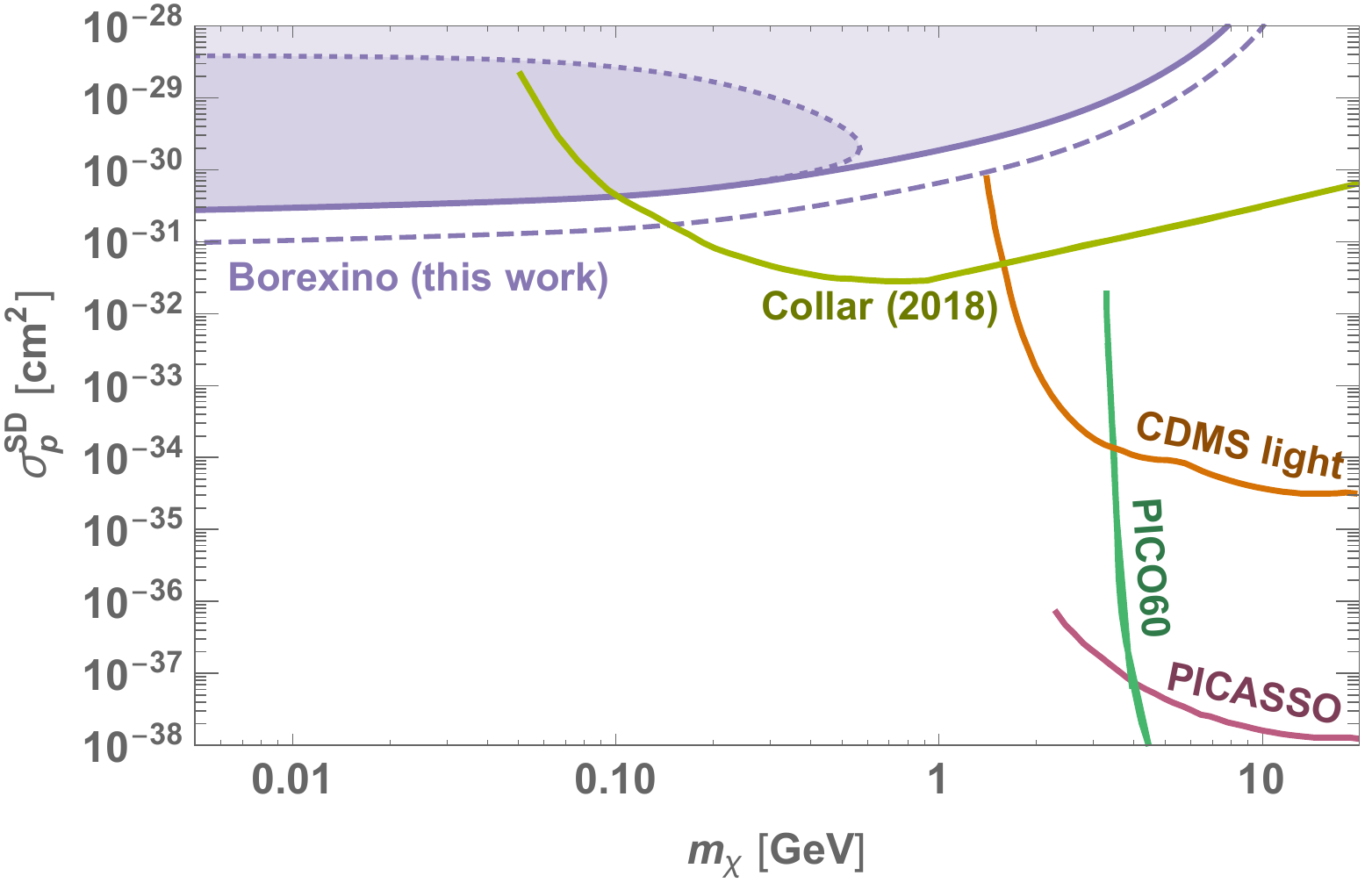}
 \caption{%
 Constraints on the spin-dependent part of the cross section imposed by Borexino. 
 Solid and dashed lines as in Fig.~\ref{constraintsSI}. Dotted lines result from adopting
 the much greater stopping power expected for spin-independent scattering
 (so this contour also applies to $\sigma_{\rm SI}$).
 For comparison we indicate limits from the direct detection experiments 
 CDMS light~\cite{Agnese:2017jvy}, PICO60~\cite{Amole:2017dex} and 
 PICASSO~\cite{Behnke:2016lsk}, as well as from 
 delayed-coincidence searches in near-surface detectors by Collar~\cite{Collar:2018ydf}.
   \label{ConstraintsSD}}
\end{figure}

Strong constraints on {\it spin-dependent scattering}, finally, can be obtained from 
proton upscattering by CRDM in neutrino detectors like Borexino~\cite{Alimonti:2000xc}.
From a search 
for events with higher energy than solar neutrino scattering~\cite{Bellini:2009jr,Bellini:2013uui}, 
we deduce that the limiting scattering rate per proton is  (see Appendix)
\begin{equation}
\Gamma_p (T_e>12.5\,{\rm MeV})< 2\times 10^{-39}\,{\rm s}^{-1}.
\end{equation}
To apply this limit, we need to convert the proton recoil energy to an {\em apparent} electron $T_e$ equivalent. 
For liquid scintillators the recoil energy of the nucleus, $T_N$, and the detected  
energy $T_e$ are related by the empirical law 
\be
 T_e(T_N)=\int_0^{T_N} \frac{d T_N}{1+k_B\langle d T_N/dx\rangle}\,,
 \label{eq:emp}
\ee
where $k_B$ is a material-dependent constant. Following the procedure 
outlined in Ref.~\cite{Dasgupta:2011wg}, and 
thus using  PSTAR tables from {\tt http://physics.nist.gov}
for $\langle d T_N/dx\rangle$, we numerically tabulate and invert Eq.~(\ref{eq:emp}) for 
pseudocumene (the scintillator used by Borexino).
The resulting constraint on spin-dependent scattering is plotted in Fig.~\ref{ConstraintsSD}.
Here the CRDM component is produced exclusively by $p-\chi$ collisions,
since $^4$He nuclei do not carry spin. 
For the mean free path in Eq.~(\ref{eq:ldef}), we assumed exclusively 
elastic scattering on nuclei as derived from spin-dependent couplings 
$\sigma_\chi=\sigma_n=\sigma_p$ to nucleons (dashed and solid lines). 
In reality, quasi-elastic scattering on nucleons would dominate for energy transfers above 
typical nuclear binding energies. While a full treatment of these processes is  
beyond the scope of this work, we indicate for comparison (dotted lines) the limits that 
would result in the extreme case of adopting a stopping power as efficient as in the case 
of spin-independent scattering, c.f.~Eq.~(\ref{eq:chiN}). For $m_\chi\lesssim0.5$\,GeV 
we thus find highly competitive limits on (both spin-independent and) spin-dependent
scattering with protons, independent of the attenuation of the CRDM flux.

\smallskip
\paragraph{Conclusions.---} We have shown that the DM-nucleon interaction cross section 
$\sigma_\chi$ necessarily generates a small but very energetic component of the DM flux, the CRDM. 
Subsequent scattering of CRDM in DM and neutrino detectors leads to novel limits on $\sigma_\chi$ 
in the $\{m_\chi, \sigma_\chi \}$ domains that previously were thought to be completely
inaccessible for direct detection. 
Our results thus complement and strengthen previous 
studies addressing the alteration of the CR spectrum, the generation of gamma rays in the 
collision of CR with DM, as well as cosmological constraints on $\chi$-nucleon 
interactions~\cite{Cyburt:2002uw,Gluscevic:2017ywp,Hooper:2018bfw,Xu:2018efh,Cappiello:2018hsu}.
All routines to calculate these new limits have been implemented in \ds~\cite{Bringmann:2018lay},
and publicly released with version 6.2 of this code~\cite{DSweb}. While our limits are generic and derived with 
a minimum set of assumptions, further refinements of the limits can be achieved within specific models, 
where $\sigma_\chi$ and its energy dependence are expressed in terms of couplings and masses 
of an underlying microscopic model. 
\phantom{\cite{Bellini:2008mr,Agostini:2017cav}}

\vfill
{\it Acknowledgments.}
We thank John Beacom, Christopher Cappiello, Juan Collar, Felix Kling and Shafi Mahdawi  
for useful discussions and input on a previous version of the manuscript.
TB wishes to warmly thank Perimeter Institute for Theoretical Physics for generous support 
and hospitality during the preparation of this manuscript. Research at Perimeter Institute is 
supported by the Government of Canada through the Department of Innovation, Science 
and Economic Development and by the Province of 
Ontario through the Ministry of Research, Innovation and Science.

\appendix

\newpage
\section{Appendix}

\paragraph*{Counting rate in deep underground neutrino detectors.---} 
We focus on Borexino (a similar analysis would apply to Kamland and SNO+), and study the  proton recoil, 
with some recoil energy $T_p$.  In the Borexino detector,  $T_p$ is quoted in terms of an {\em equivalent} 
electron recoil energy $T_e$, $T_e < T_p$. 
The main text of our Letter contains all relevant details to obtain $T_e(T_p)$. 

The most relevant for our purposes  are the ``high-energy" data by Borexino. These include studies of 
the $^8$B solar neutrino spectrum~\cite{Bellini:2008mr,Agostini:2017cav}, 
and searches of some exotic phenomena, namely violation of Pauli statistics and sterile neutrino decay~\cite{Bellini:2009jr,Bellini:2013uui}. 
The most important feature of the Borexino spectrum is a significant reduction of the radioactive backgrounds at higher $T_e$. 
For $T_e>$5 MeV these backgrounds are very small and dominated by solar $^8$B neutrinos, and above 10-12 MeV, there are ``no events" 
quoted in Ref.~\cite{Bellini:2013uui}.

We use Ref. \cite{Bellini:2013uui} to determine the total rate of proton recoil with $T_e> 12.5\,{\rm MeV}$. The constraint on the rate per 
individual proton is given by
\begin{equation}
\label{result1}
\Gamma_{p}^{\rm Borexino}(T_e > 12.5\,{\rm MeV}) = 
\frac{S_{lim}}{\epsilon\, N_p\, T} < 1.9\times 10^{-39} \, {\rm s}^{-1}
\end{equation}
In this formula, $S_{lim} =2.44$ at 90\% c.l., $T= 1.282$yr is the data-taking period, and $N_p =3.2\times 10^{31}$ 
is the total number of protons. ($N_p$ is recalculated from the number of Carbon atoms, 
$N_C = 2.37\times 10^{31}$~\cite{Bellini:2013uui}). The Borexino collaboration used an efficiency $\epsilon = 0.5$ 
for the specific search of gamma emission at $E\sim$16 MeV, but this measures the efficiency of detecting a peak. Since we are not concerned with exact energy reconstruction, we can take $\epsilon =1$ (meaning that every recoil with $T_e >12.5$ MeV would be detected). Result (\ref{result1})
is shortened in the main text as Eq. (18). 

\paragraph*{Shallow near surface detectors.---}
Shallow-site neutrino detectors, where the counting rates are much larger due to large backgrounds, nevertheless can be used as 
a useful limit when the penetration into $\sim {\rm km}$ depths is impeded by a relatively large cross section. Shallow detectors 
include MiniBooNE that have measured out-of-beam-pulse backgrounds consistent with $p+\chi \to p +\chi$ scatterings. 

To estimate the limiting counting rates we use the recent MiniBooNE dark matter search (where the proton beam is passed around the 
Be target to minimize the beam neutrino background) \cite{Aguilar-Arevalo:2017mqx}. This paper draws on the 
measurement by the same collaboration of the neutral current (NC) scattering of beam neutrinos \cite{Aguilar-Arevalo:2013nkf}.

To determine the limiting counting rate, we take the number of beam-unrelated background events \cite{Aguilar-Arevalo:2017mqx}, 
and the {\rm effective} running time $T$ given by the product of the recorded time around the beam pulse $\Delta t = 19.2 \,\mu{\rm sec}$,
times the number of bunches received, determined by the total proton-on-target for the run, $POT = 1.86\times 10^{20}$,  divided by the average number of protons per bunch  
$N_{p,{\rm bunch}} = 4\times 10^{12}$,
\begin{equation}
\Gamma_{p}^{\rm MiniBooNE}(T_p > 35\,{\rm MeV}) =
\frac{S_{\rm lim}}{\epsilon \, N_p \, \Delta t \, (POT/N_{p,{\rm bunch}})} \,.
\end{equation}
This way we get $\Delta t \times (POT/N_{p,bunch})= 893$ seconds. Using the reported beam-unrelated background, we find 
less than $S_{\rm lim} = 800$ events, i.e.~the background rate is $\sim$ Hz. 
Taking the total number of free protons in the MiniBoone inner volume to be $N_p = 5.83 \times 10^{31}$, we arrive at the limiting counting rate per proton as 
\begin{equation}
\Gamma_{p}^{\rm MiniBooNE}(T_p > 35\,{\rm MeV}) <1.5\times 10^{-32}  \, {\rm s}^{-1}.
\end{equation}
This is a conservative limit, as the (only approximately known) cosmic ray background is not subtracted from this counting rate. It is quoted as Eq. (17) in the main
text.

We note that the MiniBooNE counting rates are many orders of magnitude larger than for Borexino, but they make a difference for cross sections around $10^{-27} {\rm cm}^2$,
where energetic particles do not reach deep locations, but can still penetrate the atmosphere and $\sim 3$m of soil to the MiniBooNE detector.

\bibliography{Ref}

\end{document}